\documentclass[aip,graphicx]{revtex4-1}

\usepackage{amsmath}
\usepackage{comment}
\usepackage{amsxtra}
\usepackage{amsmath}
\usepackage{amssymb}
\usepackage{mathrsfs}
\usepackage{graphicx}
\usepackage{natbib}
\usepackage{color}
\usepackage{amsmath}
\usepackage{setspace}
\usepackage{array}
\def\strutdepth{\dp\strutbox}
\def\nw#1{\strut\vadjust{\kern-\strutdepth\vtop to0pt{\vss\hbox
to\hsize
{\hskip\hsize\hskip5pt$\leftarrow$\hss\strut}}}{\em #1}}
\definecolor{gris}{gray}{0.60}

\begin{document}


\title{Paper waves in the wind}

\author{Pan Jia}
\email[]{pan.jia@espci.fr}
\author{Bruno Andreotti}
\author{Philippe Claudin}
\affiliation{Laboratoire de Physique et M\'ecanique des Milieux H\'et\'erog\`enes\\
PMMH UMR 7636 ESPCI -- CNRS -- Univ.~Paris-Diderot -- Univ.~P.M.~Curie,\\
10 rue Vauquelin, 75005 Paris Cedex 05, France}

\date{\today}

\begin{abstract}
A flexible sheet clamped at both ends and submitted to a permanent wind is unstable and propagates waves. Here, we experimentally study the selection of frequency and wavenumber as a function of the wind velocity. These quantities obey simple scaling laws, which are analytically derived from a linear stability analysis of the problem, and which also involve a gravity-induced velocity scale. This approach allows us to collapse data obtained with sheets whose flexible rigidity is varied by two orders of magnitude. This principle may be applied in the future for energy harvesting.
\end{abstract}

\maketitle

\section{Introduction}
\label{Introduction}

Problems involving the coupling between a fluid and a structure or a deformable body have attracted a lot of attention in fundamental science and engineering\cite{Book-Paidoussis2004}. Besides traditional applications like the design of aircraft\cite{Book-Anderson1997}, automobiles\cite{AnnulReviewFM_Katz2006} and bridges\cite{JWEIA_Toshio2003}, fluid-structure interaction has also recently been considered for medical treatments, such as in the analysis of aneurysms in large arteries\cite{CS-Gerbeau2005} and artificial heart valves\cite{JBM-Lim2003}. Among the various topics in this field,  a classical setup is the case of a cantilevered flexible sheet lying in an axial flow, attached on the up-stream side and freely flapping at the down-stream end, usually termed as the flag instability. This problem is related to paper industry\cite{JFS-Watanabe2002a,JFS-Watanabe2002b} and airfoil flutter\cite{Book-Bisplinghoff1983}, as well as to biological situations including snoring\cite{JFS-Huang1995} and the motion of swimming or flying animals\cite{JFM-Lighthill1960,Nature-Huber2000,Science-Liao2003,Science-Muller2003,PNAS-Ramananarivo2011,Book-Shyy2013} -- see a recent review by Shelley \& Zhang\cite{AnnRevFM-Shelly2011} and references therein. A related but different situation is that of a flexible or compliant material clamped at both ends -- thus avoiding the flapping phenomenon and subsequent vortex shedding from the trailing edge -- which develops traveling waves when submitted to a flow\cite{JSV-Hansen1974,JSV-Kornecki1976,JSV-Hansen1980,JFM-Kim2014}. Finally, the interaction between a fluid and a structure is also relevant in the context of geological fluid mechanics\cite{JFM-Huppert1986}. Famous examples are the dynamics of meanders\cite{JFM-Seminara2006} or the formation of sand ripples and dunes at the surface of an erodible bed\cite{AnnRevFM-Charru2013}.

The flag instability results from the competition between the destabilising effect of the aerodynamic pressure, which, by virtue of Bernouilli's principle, is lowered above bumps and increased in troughs, and the stabilising effect of the bending rigidity of the solid, which tends to keep the sheet flat. This explanation was already reported by Rayleigh\cite{Rayleigh1879}, who found that a flag of infinite span and infinite length is always unstable. When considering a flag of finite dimensions, this problem becomes more difficult and depends on its aspect ratio, defined as the ratio of the flag span to the flag length. The slender body approach and the airfoil theory have been respectively employed to implement the calculations for asymptotically small\cite{JFM-Lighthill1960,JFS-Lemaitre2005}and large\cite{JSV-Kornecki1976,JFS-Huang1995,JFS-Watanabe2002b,JApplMech-Guo2000} aspect ratios. A recent unified model by Eloy et al.\cite{JFS-Eloy2007,JFM-Eloy2008} has considered the intermediate case. Experimental studies have been carried out in wind tunnels\cite{JPSJapan-Teneda1968,JFS-Huang1995,JFS-Watanabe2002b}, in water flumes\cite{PRL-Shelley2005} and even in flowing soap films\cite{Nature-Zhang2000}. The first experiments of Taneda were performed in a vertical wind tunnel with flags of different materials (silk, flannel, canvas, muslin) and shapes (triangles, rectangles), and it is reported that the flags do not flap in slow flows due to the stabilizing effects of both viscosity and gravity\cite{JPSJapan-Teneda1968}. Later, Datta \& Gottenberg conducted experiments with long ribbons hanging vertically in downward flows, and the critical flow velocity for the onset of flapping was studied as a function of the length, width and thickness of the ribbons\cite{JApplMech-Datta1975}. This small aspect ratio regime has recently been experimentally revisited by Lemaitre\cite{JFS-Lemaitre2005}. Experiments for larger and intermediate aspect ratios have also been reported by Huang\cite{JFS-Huang1995}, Watanabe et al\cite{JFS-Watanabe2002b}, Yamaguchi et al\cite{JFE-Yamaguchi2000} and Eloy et al\cite{JFM-Eloy2008}. In these experiments, the critical velocity is systematically found higher than theoretical predictions\cite{PNAS-Argentina2005} and the origin of this discrepancy has recently been related to inherent planarity defects\cite{JFM-Eloy2012}. In a more recent experiment, Kim et al. have investigated the occurrence of the flapping of an inverted flag, with a free leading edge and a fixed trailing edge\cite{JFM-Kim2013}. Various numerical approaches have also been used to tackle the different aspects of this problem\cite{JFS-Watanabe2002a,PRL-Alben2008,JFM-Michelin2008,JSV-Tang2008}.

Here we report experiments of wind-generated waves on a flexible sheet clamped at both ends. These elastic surface waves are induced by the same pressure-related instability mechanism as in the case of the flag flapping, but their dynamics and length/time scale selection are different. We measure the frequency and the wavenumber of the waves varying the wind velocity and for different materials (paper and plastic sheets). We show that they obey simple scaling laws, in agreement with a linear stability analysis based on a simple hydrodynamic assumption and on the Euler-Bernouilli beam theory for the sheet. The paper is organized as follows. The experimental setup and the data processing techniques are described in section 2; In section 3, the linear instability analysis is carried out. Finally, experimental results are compared with the theoretical predictions in section 4.

\section{Experiments}
\label{experiments}

\subsection{Experimental Setup}
\begin{figure}[t]
\centering
\includegraphics{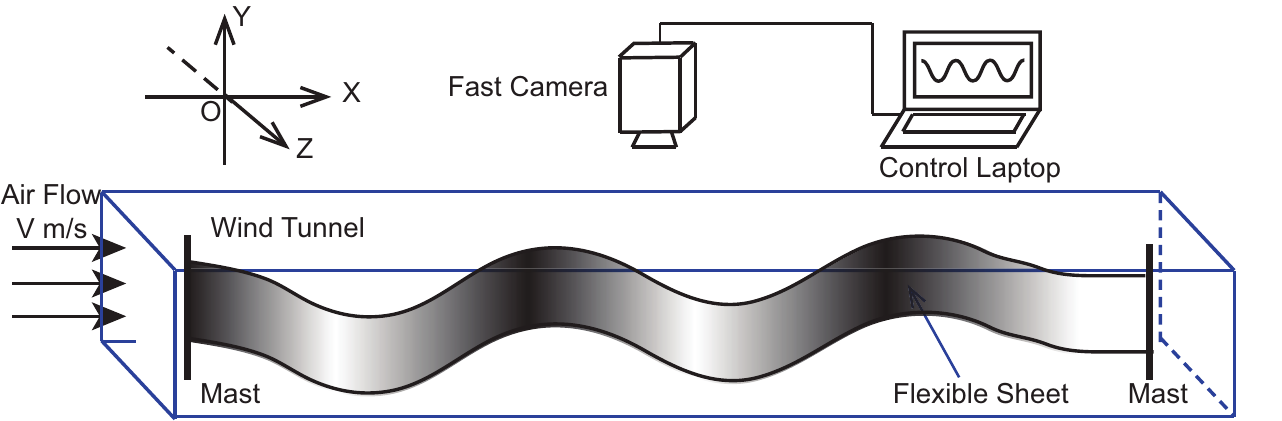}
\caption{Sketch of the experimental setup. $V$ is the air flow velocity, measured at a fixed position at the outlet of the tunnel. The distance between the two masts is $L_0=1.5$~m. The square section of the tunnel is $0.15\;\textrm{m} \times 0.15 \;\textrm{m}$. The sheet is $4$~cm wide and its total length is denoted as $L = L_0 + \Delta L$. $\Delta L$ has been varied from $4$ to $12$~cm.}
\label{fig:setup}
\end{figure}

As schematically illustrated in Fig.~\ref{fig:setup}, the experiments are conducted in a wind tunnel with a square cross section. We denote as $x$, $y$ and $z$ the longitudinal, transverse vertical and horizontal axes, respectively. The wind flow is induced along $x$ by imposing the pressure at the inlet. The wind velocity $V$ is monitored with an anemometer (Testo 405-V1) at a fixed position at the exit of the tunnel. A flexible sheet of width $\simeq 4$~cm and of length $L$ is placed at the centre of the tunnel and clamped at both ends on fixed masts. We denote as $L_0$ the distance between the two masts and define $\Delta L = L - L_0$ (different values of $\Delta L$ have been used, see below). The reference coordinate $x=0$ is chosen at the inlet mast. The air flow is uniformly injected on both faces of the sheet, in order to avoid the formation of vortices. A transverse orientation of the sheet along $y$ and $z$ have both been tested, showing qualitatively similar behaviors (quantitative differences exist, due to gravity, see Section~\ref{results}), and in what follows all data correspond to a vertical orientation. The experiments are conducted either with paper or plastic (bi-oriented polypropylene) sheets of different thicknesses. The values of the relevant physical characteristics of these materials are given in Tab.~\ref{tab:MaterialParameterValues}, namely the mass per unit surface $m$ and the bending rigidity $D$. This last quantity was measured by determining the deflection under gravity of horizontally clamped strips of various lengths. Interestingly, $D$ changes by two orders of magnitude from the thin plastic sheet to the paper, which allows us to investigate the instability over a large range of parameters. A fast camera (Phantom Miro M340) is employed to record the motion of the sheet through the transparent ceiling of the tunnel. The camera is operated with a spatial resolution of $2560\;\textrm{pixel} \times 320\;\textrm{pixel}$, at $1.5$~pixel/mm, and an exposure time of $400\;\mu$s. The sample rate is typically between $1000$~Hz and $2500$~Hz during the measurements.

\begin{table}[h!]
\centering
\caption{Values of the mass per unit surface $m$ and the bending rigidity $D$ of the three different sheets used in the experiments.}
\label{tab:MaterialParameterValues}
\begin{tabular}{p{3cm}<{\centering}p{3cm}<{\centering}p{3cm}<{\centering}} 
\hline
Material	&  $m$ 				& $D$				\\
		& (kg/m$^{2}$)			& (Nm)				\\
\hline
Paper	& $2.5\times 10^{-1}$	& $7.6\times 10^{-3}$	\\
Plastic	& $8.6\times 10^{-2}$	& $1.5\times 10^{-4}$	\\
Plastic	& $4.9\times 10^{-2}$	& $4.2\times 10^{-5}$	\\
\hline
\end{tabular}
\end{table}

\begin{figure}[t]
\centering
\includegraphics{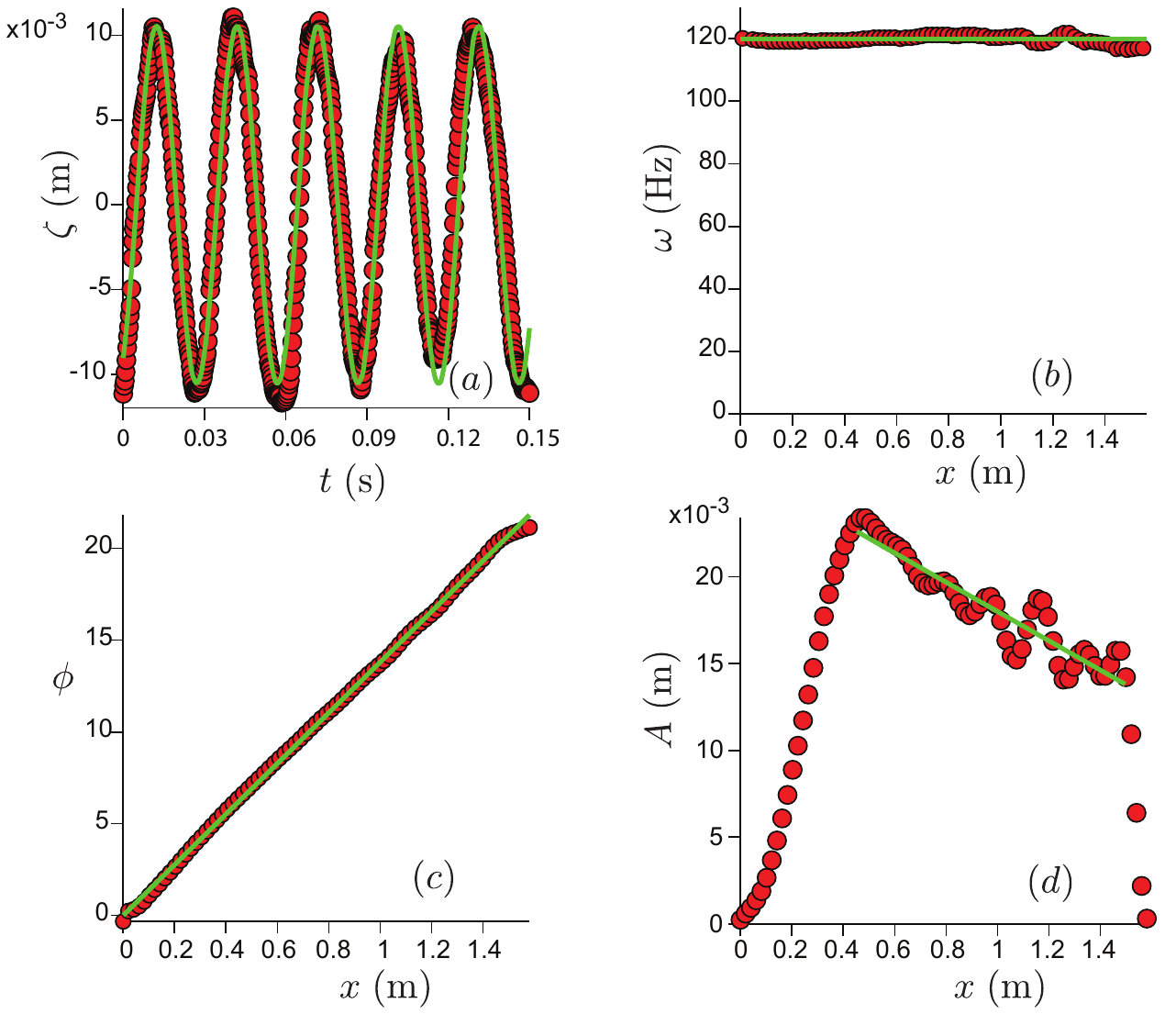}
\caption{(Color online) Typical experimental measurements and data processing. (a) Temporal variation of the deflection $\zeta$ of the sheet at a given location (here, $x=0.235$~m). Solid line: Fourier fit of the form $\zeta(x,t) = A \cos (\omega t - \phi)$.
(b) Spatial variation of the angular frequency $\omega$ along $x$. Solid line: fit by a constant.
(c) Spatial variation of the phase $\phi$ along $x$. Solid line: linear fit of the form $\phi = k x$.
(d) Spatial variation of amplitude $A$ along $x$. Solid line: linear fit to estimate the slope $\simeq 10^{-2}$.
All these data correspond to the paper sheet, at $V = 8.6$~m/s and with $\Delta L = 4$~cm. Data for the plastic sheets or for other control parameters look very similar. In panels (b-d), $8$ time series corresponding to $2000$ frames at $1000$~Hz ($2$~s in total) have been averaged.}
\label{BasicData}
\end{figure}

\subsection{Experimental Data}
The wind flow generates waves on the sheet, that propagate downstream. The high-frequency movies allow us to follow in detail the kinematics of these waves. In the case of the paper sheet, they are mostly transverse to the wind, i.e. the sheet is not twisted. However, due to a smaller flexibility, the plastic sheets exhibit three-dimensional motions for the stronger winds. For a given wind velocity, the experimental data are obtained as follows. The borders of the sheet are detected from the image sequences by maximizing the correlation with an analyzing wavelet. The average profile $\zeta(x,t)$ representing the two-dimensional shape of the sheet is determined with a sub-millimetric resolution. The typical time variation of $\zeta$ at a fixed value of $x$ is displayed in Fig.~\ref{BasicData}a, and shows harmonic oscillations: for a given $x$, $\zeta(x,t)$ is well represented by the function $A \cos (\omega t - \phi)$, where $A$ is an amplitude; $\omega$ is an angular frequency and $\phi$ is a phase. All these three quantities \textit{a priori} depend on $x$ (and $V$). However, as shown in Fig.~\ref{BasicData}b, $\omega$ turns out to be constant all along the sheet and the phase is linearly related to space $\phi = k x$ (Fig.~\ref{BasicData}c), corresponding to a constant wave propagation velocity $c=\omega/k$. $k$ is the wavenumber of the waves, and $\lambda = 2\pi/k$ is the wavelength. The behavior of $A(x)$ shows several regimes (Fig.~\ref{BasicData}d). $A$ vanishes at both ends of the sheet, as it should because of the clamping. In between, it first increases rapidly, then slowly decreases in a more noisy way over most of the tunnel, and finally quickly drops at the very end. The first part can be associated with the spatial development of the instability. As shown in section~\ref{results}, the typical value of the amplitude in the second regime is dictated by the geometrical constraint that relates $A$ to $k$ and $\Delta L/L_0$.

We have conducted experiments similar to that corresponding to Fig.~\ref{BasicData}, systematically varying the air flow velocity and the total length of the sheet $L$, for all three types of sheet. We now mostly focus on the frequency and the wavenumber of the waves, for which a comparison with an analytical theory is possible.

\section{Linear stability analysis}
\label{modelling}

The purpose of this section is to provide a brief but self-contained summary of the theoretical framework within which we analyze our experimental data. The theoretical description of the flow over a flexible sheet has been treated in a general way several decades ago, as e.g. summarized by Paidoussis \cite{Book-Paidoussis2004}, see chapter ~10 and references therein. Here, we restrict this analysis to a two-dimensional linear perturbation theory. Furthermore, we hypothesize that the air flow can be decomposed into a turbulent inner boundary layer and an outer laminar flow which can be described as an incompressible perfect fluid. As we only need the pressure field, which is almost constant across the inner layer, we will simply describe the outer layer. Under these simplifying assumptions, we are able to derive analytical scaling laws for the frequency and the wavenumber of the most unstable mode in the asymptotic limit of either very flexible or very rigid sheets, and which were not available in the literature.

\subsection{Governing equations}
We consider a flexible sheet of infinite span and length submitted to an air flow along the $x$-axis. For later rescalings, we denote as $V$ the characteristic velocity of the wind, i.e. the average air velocity at a given and fixed altitude $z_w$ (in the experiment $z_w$ is on the order of a few cm). Assuming that the motion of the sheet is independent of the coordinate $y$, we denote $\zeta(x,t)$ as its deflection with respect to the reference line $z=0$. In the limit of small deflections with respect to a flat reference state, the sheet obeys the linearized Euler-Bernoulli beam equation:
\begin{equation}
m\frac{\partial^2 \zeta}{\partial t^2} + D \frac{\partial^4 \zeta}{\partial x^4} + \delta p = 0,
\label{Euler-Bernoulli}
\end{equation}
where $\delta p$ the air pressure jump across the sheet.

Neglecting viscous stress in the outer layer and assuming incompressibility (recall that velocities are on the order of a few m/s, i.e. corresponding to very low Mach numbers), mass and momentum conservations for the flow field are therefore expressed by Euler equations:
\begin{eqnarray}
\nabla \cdot \textbf{u} & = & 0,
\label{Euler_mass}\\
\frac{\partial \textbf{u}} {\partial t}+\left(\textbf{u} \cdot \nabla \right) \textbf{u} & = & -\frac{1}{\rho} \nabla p,
\label{Euler_momentum}
\end{eqnarray}
where $\textbf{u}$ and $p$ are the velocity and pressure fields, and $\rho$ is the air density. Finally, the fluid velocity on the sheet should be equal to the sheet velocity, in order to ensure the impermeability of the sheet:
\begin{equation}
\textbf{u} (z=\zeta) \cdot \textbf{n} =\frac{\textrm{d}\zeta}{\textrm{d}t}\, ,
\label{ImpermeabilityCondition}
\end{equation}
where $\textbf{n}$ is the unit vector normal to the sheet. Equations (\ref{Euler_mass}-\ref{ImpermeabilityCondition}) must be consistently linearized in the limit of small sheet deflections, and together with (\ref{Euler-Bernoulli}) they form a closed set that we analyze in the next sub-section.

\subsection{Linearized problem}
We consider that the sheet is long enough to ignore the influence of boundaries. Considering an infinite sheet, invariant along $x$, the normal modes are sinusoidal and characterized by the complex frequency $\Omega$. We therefore look for solutions of the linearized problem of the form:
\begin{eqnarray}
\zeta &=& A e^{ikx - i\Omega t},
\label{defzeta} \\
u_x & = & V + \mathcal{U} V kA e^{- i\Omega t + ikx - \kappa z},
\label{NormalModesU} \\
u_z & = & \mathcal{W} V kA e^{- i\Omega t + ikx - \kappa z},
\label{NormalModesW} \\
p & = & P_0 + \mathcal{P} \rho V^2 kA e^{- i\Omega t + ikx - \kappa z},
\label{NormalModesP}
\end{eqnarray}
where $P_0$ is the reference pressure. $\Omega$ will be later decomposed into real and imaginary parts as $\Omega = \omega + i \sigma$, where $\sigma$ is the temporal growth rate of the perturbation. $\kappa$ is the spatial decay rate along the $z$-axis. Each of these expressions consists in the sum of a zeroth order term corresponding to the flat reference state (it is then $0$ for $\zeta$ and $u_z$), and a first order term in $e^{- i\Omega t + ikx - \kappa z}$. $A$ sets the amplitude of the sheet deflection, and, as it should in the framework of a linear analysis, it will factor out of all results when solving the governing equations in the asymptotic limit $kA \ll 1$. $V$ sets the dimensionful reference for the velocities, and $\rho V^2$ does that for the pressure. The unknowns are thus the dimensionless quantities $\mathcal{U}$, $\mathcal{W}$ and $\mathcal{P}$ as well as $\kappa$, and they are to be determined by the above governing equations.

Plugging (\ref{defzeta}-\ref{NormalModesP}) into (\ref{Euler-Bernoulli}-\ref{ImpermeabilityCondition}), and treating the first order in $kA$, we first find
\begin{equation}
\kappa^2 = k^2,
\label{kappa}
\end{equation}
from which we set $\kappa=k$ (resp. $\kappa = -k$) for the fields in the region $z>0$ (resp. $z<0$), in order for the perturbation to decay away from the sheet. The three other unknowns are found as:
\begin{eqnarray}
\mathcal{U} & = & \pm \frac{kV-\Omega}{kV},
\label{mathcalU} \\
\mathcal{W}  & = & i \, \frac{kV-\Omega}{kV},
\label{mathcalW} \\
\mathcal{P} & = & \mp \frac{(kV-\Omega)^2}{(kV)^2},
\label{mathcalP}
\end{eqnarray}
where the $\pm$ sign corresponds to the positive/negative region (in $z$). For our purpose, the most important quantity is $\delta p$ the pressure jump across the sheet,  which can be written as:
\begin{equation}
\delta p = p(z \to 0^+) - p(z \to 0^-) = - 2\rho \frac{(kV-\Omega)^2}{k} \zeta.
\label{PressureDrop}
\end{equation}
This allows us to express the dispersion relation from Eq.~\ref{Euler-Bernoulli} as:
\begin{equation}
m\Omega^2 - Dk^4 + 2\rho \frac{(kV-\Omega)^2}{k}=0,
\label{DispRel_Dim}
\end{equation}
whose properties are studied in the next sub-section. Note again that this equation is a simpler form of the dispersion relation derived in reference 1 (Chapter 10), where the distance to the bottom wall has been sent to infinity and the viscous drag, the spring stiffness of the elastic foundation and the plate tension have been set to zero.

\subsection{Dispersion relation}
The above equation can be made dimensionless by setting $\bar{\Omega}=\frac{ m \Omega}{\rho V}$ for the frequency, $\bar{k}=\frac{mk}{\rho}$ for the wavenumber, and $\bar{D}=\frac{\rho^2 D}{m^3V^2}$ for the bending rigidity. With these rescaled variables, Eq.~\ref{DispRel_Dim} can be written as
\begin{equation}
\bar{\Omega}^2-\bar{D} \bar{k}^4+\frac{2}{\bar{k}}(\bar{k}-\bar{\Omega})^2=0,
\label{DispRel_aDim}
\end{equation}
and $\bar{D}$ is its only parameter. Because it is quadratic in $\bar{\Omega}$, (\ref{DispRel_aDim}) can easily be solved as:
\begin{equation}
\bar{\Omega}=\frac{2\bar{k}}{\bar{k}+2} \pm \sqrt{\frac{-2\bar{k}^3+2\bar{D}\bar{k}^5+\bar{D}\bar{k}^6}{(\bar{k}+2)^2}}.
\label{OmegaPM}
\end{equation}
Due to the square root in (\ref{OmegaPM}), we can identify a cutoff wave number $\bar{k}_c$, solution of
\begin{equation}
-2+2\bar{D}\bar{k}_c^2+\bar{D}\bar{k}_c^3=0.
\label{Equakc}
\end{equation}
Above $\bar{k}_c$, the complex frequency $\bar{\Omega}$ of any perturbation is in fact real (the expression below the square root is positive), corresponding to a wave which propagates without growth nor decay ($\sigma=0$). On the other hand, any perturbation with a wavenumber below $\bar{k}_c$ has an angular frequency and a growth rate given by:
\begin{eqnarray}
\bar{\omega} & = & \frac{2\bar{k}}{\bar{k}+2} \, ,
\label{omega} \\
\bar{\sigma} & = & \pm \sqrt{\frac{2\bar{k}^3-2\bar{D}\bar{k}^5-\bar{D}\bar{k}^6}{(\bar{k}+2)^2}} \, .
\label{sigma}
\end{eqnarray}

\begin{figure}[t]
\centering
\includegraphics{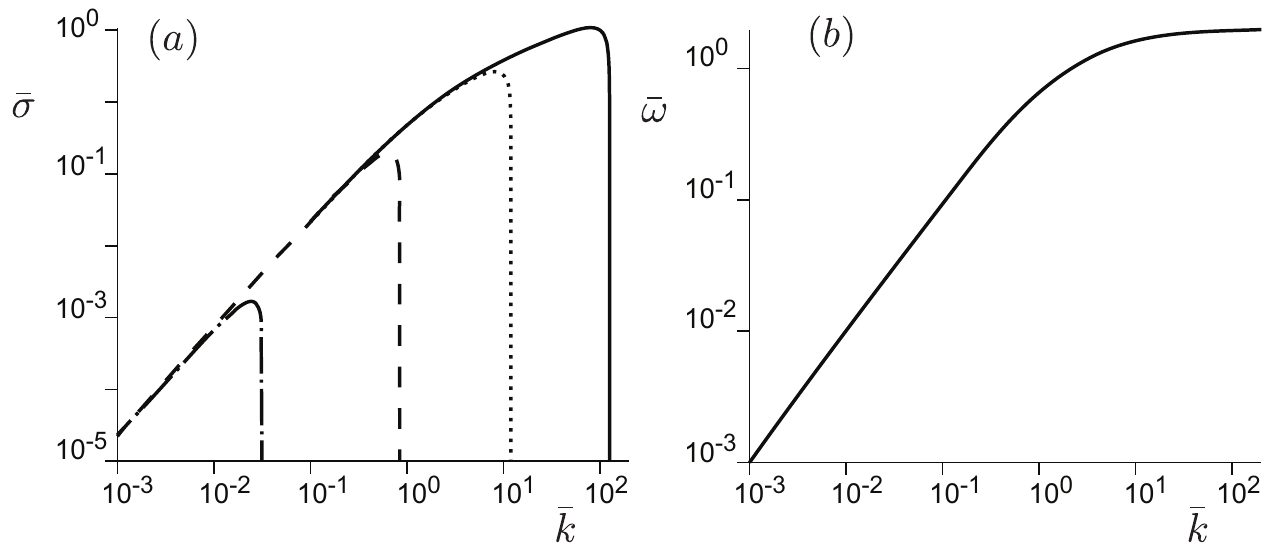}
\caption{(a) Dimensionless growth rate $\bar{\sigma}$ (positive branches only) as a function as the dimensionless wavenumber, for $\bar{k} < \bar{k}_c$, Eq.~\ref{sigma}. The different lines correspond to different values of the bending rigidity $\bar{D}$. Dotted-dashed line: $\bar{D}=10^3$, $\bar{k}_c \simeq 3.1 \, 10^{-2}$. Dashed line: $\bar{D}=10^0$, $\bar{k}_c \simeq 8.4 \, 10^{-1}$. Dotted line: $\bar{D}=10^{-3}$, $\bar{k}_c \simeq 1.2 \, 10^{1}$. Solid line: $\bar{D}=10^{-6}$, $\bar{k}_c \simeq 1.3 \, 10^{2}$. (b) Dimensionless angular frequency $\bar{\omega}$ as a function as of $\bar{k}$, Eq.~\ref{omega}. Note: for $\bar{k} < \bar{k}_c$, $\bar{\omega}$ is independent of $\bar{D}$.}
\label{DispRelFig}
\end{figure}

We display $\bar{\sigma}$ (positive branch) and $\bar{\omega}$ as functions of $\bar{k}$ in Fig.~\ref{DispRelFig}. All wavenumbers between $0$ and $k_c$ are unstable. In between, $\bar{\sigma}$ shows a maximum, corresponding to the most unstable wavenumber $\bar{k}_m$. It is the solution of $\textrm{d}\bar{\sigma}/\textrm{d}\bar{k}=0$, which gives:
\begin{equation}
2 \bar{D} \bar{k}_m^4+9 \bar{D} \bar{k}_m^3+10 \bar{D} \bar{k}_m^2-\bar{k}_m-6=0.
\label{Equakm}
\end{equation}
Both $\bar{k}_m$ and $\bar{k}_c$ depend on the rigidity of the sheet, and they are larger for smaller $\bar{D}$. In the unstable range of $\bar{k}$, $\bar{\omega}$ is independent of $\bar{D}$. It increases linearly with $\bar{k}$, and eventually saturates to the value $2$ when $\bar{k}$ reaches values on the order of unity. Interestingly, this saturated regime corresponds to waves with vanishing phase ($\omega/k$) and group ($d\omega/dk$) velocities. Beyond $\bar{k}_c$, $\bar{\sigma}=0$ and $\bar{\omega}$ enters another regime (not shown in Fig.~\ref{DispRelFig}b) where it asymptotically varies like the square of the wavenumber.

In this temporal stability analysis, both phase and group velocities are found positive, which may suggest a convective instability, as is often the case for instabilities generating propagative waves. However, performing the spatial stability analysis of these equations, one finds unstable modes with both positive and negative group velocities, suggesting an absolute instability. Previous theoretical analyses of this issue \cite{AnnulReviewFM_Huerre1990,JSV_deLangre2002} (and references therein) have shown that the instability is absolute below a threshold in $\bar D$ around $10^2$, i.e. at small enough $D$ or large enough $V$. This therefore justifies the temporal analysis performed here.

\begin{figure}[t]
\centering
\includegraphics{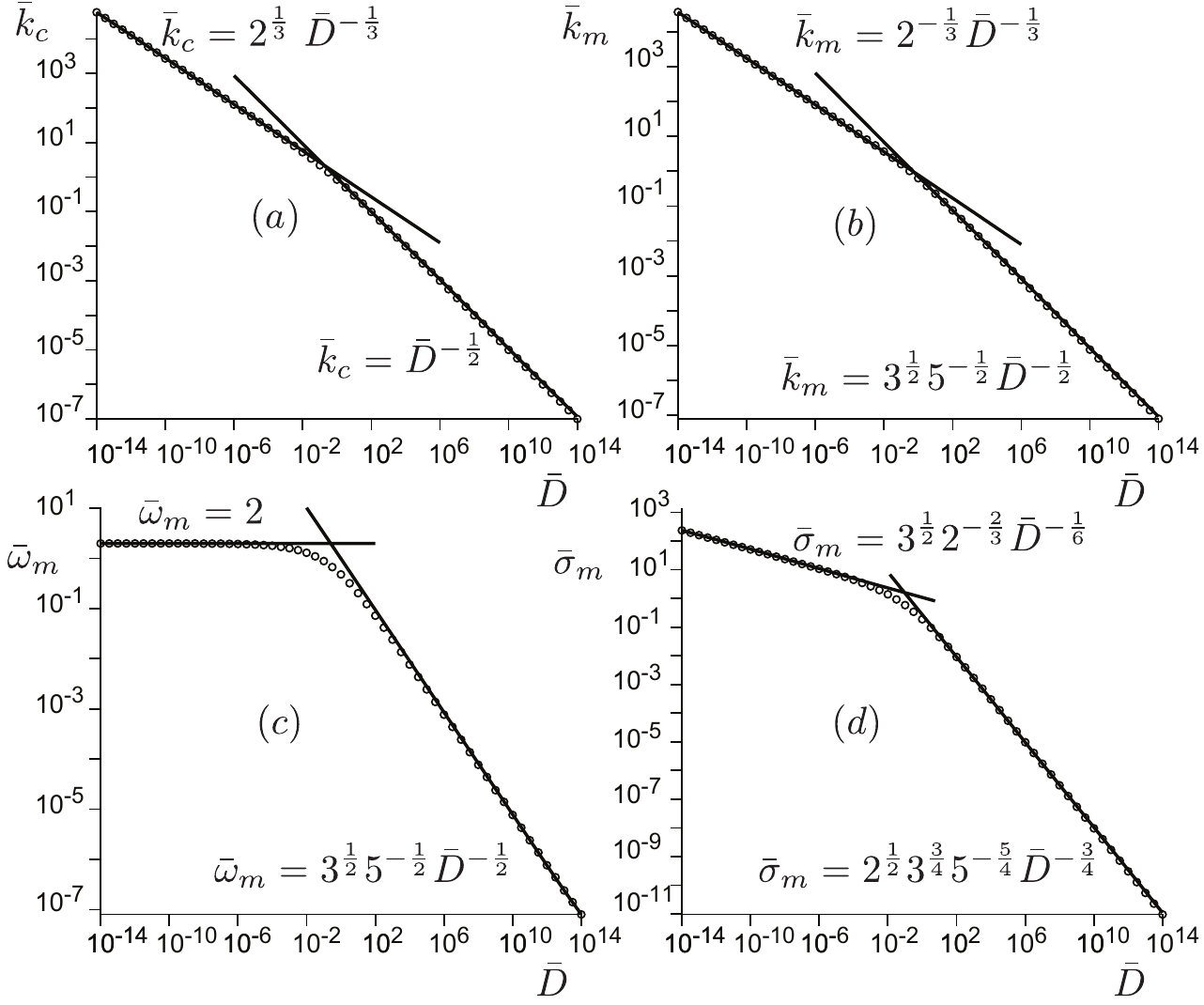}
\caption{Cut-off wavenumber $\bar{k}_c$ (a), maximum wavenumber $\bar{k}_m$ (b), maximum angular frequency $\bar{\omega}_m$ (c) and maximum temporal growth rate $\bar{\sigma}_m$ (d) as functions of the bending rigidity $\bar{D}$. Circles: data obtained numerically from the equations. Solid lines: asymptotic expressions (\ref{kckmDto0}-\ref{omegamsigmaDtoInfty}).}
\label{FigScalingsTheo}
\end{figure}

\subsection{Asymptotic analysis and scaling laws}
Scaling laws for the characteristics of the most unstable mode ($\bar{k}_m$, $\bar{\omega}_m$, $\bar{\sigma}_m$) as well as for the cut-off wavenumber $\bar{k}_c$ can be analytically derived in the limits of asymptotically small and large $\bar{D}$. $\bar{k}_c$ and $\bar{k}_m$ are calculated from (\ref{Equakc}) and (\ref{Equakm}), respectively. $\bar{\omega}_m$ and $\bar{\sigma}_m$ are obtained by introducing $\bar{k}_m$ into Eqs.~\ref{omega},\ref{sigma}. Expanding these equations in the limit $\bar{D} \to 0$ or $\bar{D} \to \infty$, one obtains for the wavenumbers:
\begin{eqnarray}
\bar{k}_c \sim \left ( \frac{2}{\bar{D}} \right )^{1/3}, \quad \bar{k}_m \sim \left ( \frac{1}{2\bar{D}} \right )^{1/3} \qquad & \textrm{for} & \qquad \bar{D} \to 0,
\label{kckmDto0}\\
\bar{k}_c \sim \left ( \frac{1}{\bar{D}} \right )^{1/2}, \quad \bar{k}_m \sim \left ( \frac{2}{5\bar{D}} \right )^{1/2} \qquad & \textrm{for} & \qquad \bar{D} \to \infty.
\label{kckmDtoInfty}
\end{eqnarray}
Similarly, the angular frequencies and growth rates scale as:
\begin{eqnarray}
\bar{\omega}_m \sim 2, \quad \bar{\sigma}_m \sim \left ( \frac{27}{16 \bar{D}} \right )^{1/6} \qquad & \textrm{for} & \qquad \bar{D} \to 0,
\label{omegamsigmaDto0}\\
\bar{\omega}_m \sim \left ( \frac{3}{5\bar{D}} \right )^{1/2}, \quad \bar{\sigma}_m \sim \left ( \frac{2}{5} \right )^{1/2} \! \left ( \frac{3}{5\bar{D}} \right )^{3/4} \qquad & \textrm{for} & \qquad \bar{D} \to \infty.
\label{omegamsigmaDtoInfty}
\end{eqnarray}
The variations of $\bar{k}_c$, $\bar{k}_m$, $\bar{\omega}_m$ and $\bar{\sigma}_m$ with $\bar{D}$ are displayed in Fig.~\ref{FigScalingsTheo}, showing a very good agreement between the numerical solution of the equations and this asymptotic analysis.

\section{Results and Discussions}
\label{results}

\subsection{Selection of angular frequency and wavenumber}
Considering the experimental parameters (see Tab.~\ref{tab:MaterialParameterValues} and typical values in Section~\ref{experiments}), the dimensionless rigidity lies in the range $10^{-3}$ -- $10^{-2}$. For the analysis of the experimental data, we shall then make use of the scaling laws (\ref{kckmDto0}) and (\ref{omegamsigmaDto0}) obtained in the limit of small $\bar{D}$. Introducing back physical dimensions in these expressions, we obtain
\begin{eqnarray}
\omega_m & \sim & \frac{2\rho}{m} V,
\label{scalingomegamV} \\
k_m & \sim & \left ( \frac{\rho}{2D} \right )^{1/3} V^{2/3}.
\label{scalingkmV}
\end{eqnarray}
The selected angular frequency purely results from the balance between dynamic pressure and inertia. The selected wavenumber results from the balance between dynamic pressure and elasticity. It is interesting to compare the phase velocity $\omega_m/k_m$ with that of the elastic waves in the absence of wind flow. In the latter case, Eq.~\ref{Euler-Bernoulli} tells us that the dispersion relation is simply $\omega = \sqrt{D/m} k^2$, which corresponds to a velocity, evaluated at the most unstable wavenumber, $\sqrt{D/m} k_m$. This scales as $k_m$ with $V^{2/3}$, whereas $\omega_m/k_m$ is here predicted to be proportional to $V^{1/3}$. Our main goal is the experimental verification of these scaling laws of $\omega_m$ and $k_m$ with the wind velocity $V$.

The influence of gravity $g$ is not accounted for in the theory. Computing the dimensionless ratio $mgh^3/D$, where $h$ is the sheet thickness, which compares the gravity-induced stress and the stress due to elastic bending, we can see that gravity is clearly sub-dominant: this number is typically in the range $10^{-9}$--$10^{-8}$. However, gravity does break the up-down symmetry by slightly twisting the sheet, and, importantly, it sets a velocity scale $v_c$ that breaks the predicted scale-free power laws. Balancing the dynamic pressure $\rho v_c^2$ with the weight of the sheet $mg$ per unit surface, we get the characteristic velocity:
\begin{equation}
v_c = \sqrt{m g/\rho}.
\label{defvc}
\end{equation}
Assuming that the observed waves correspond to the most unstable mode, we therefore expect a data collapse when plotting data in the following way:
\begin{eqnarray}
\frac{1}{2} \sqrt{\frac{m}{\rho g}} \, \omega & \sim & \frac{V}{v_c} \, ,
\label{expscalingomegaV} \\
\sqrt[3]{\frac{2D}{m g}} \, k & \sim & \left ( \frac{V}{v_c} \right )^{2/3}.
\label{expscalingkV}
\end{eqnarray}
We emphasize that $v_c$ is not an adjustable parameter whose value would depend on the experimental setting, but is part of the theoretical analysis.

\begin{figure}[t]
\centering
\includegraphics{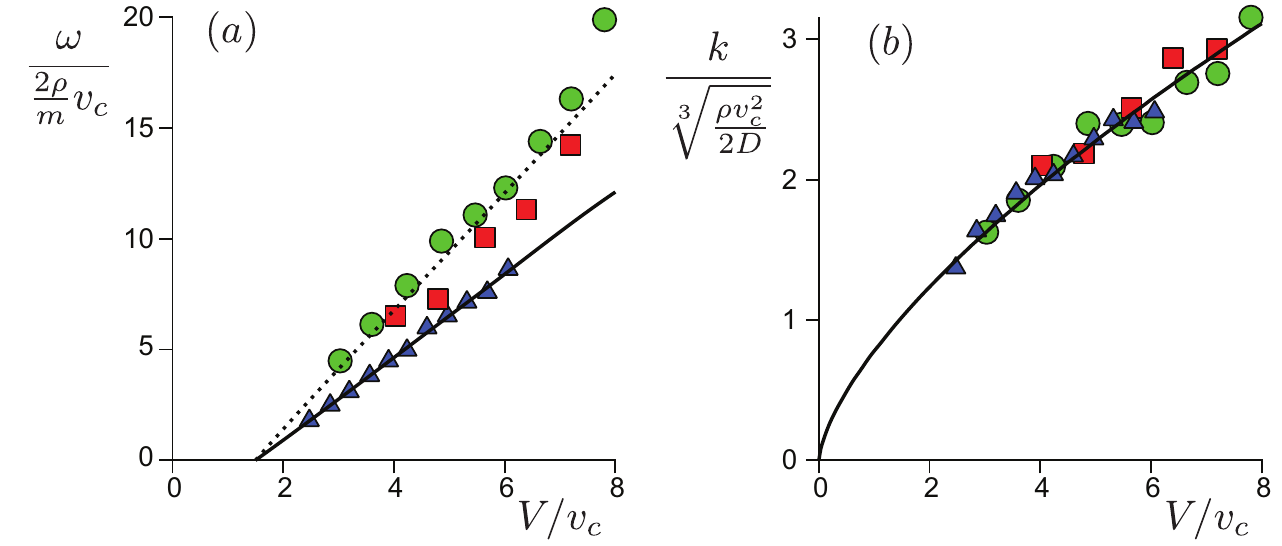}
\caption{(Color online) (a) Experimental values of the angular frequency as a function of the rescaled wind velocity. The best linear fit for the paper sheet (triangles) is shown in solid line (slope $1.9$). The dotted line is the best linear fit (slope $2.6$) for the plastic sheets (circles for the thick sheet and squares for the thin one). (b) Experimental values of the  wavenumber as a function of the rescaled wind velocity. The best fit by a 2/3-power law gives a multiplicative factor $0.78$.
All these data are for $\Delta L = 4$~cm.}
\label{ExpScalingsomegamkm}
\end{figure}

The experimental measurements of the angular frequency and the wavenumber, rescaled as proposed above, are shown in Fig.~\ref{ExpScalingsomegamkm}. The data collapse is effectively pretty good, especially if one keeps in mind that the dimensional wavenumber typically vary by a factor of $6$ and $\omega$ by a factor of $10$ at a given wind velocity from the paper to the thin plastic sheet, for which $D$ changes by two orders of magnitude (Tab.~\ref{tab:MaterialParameterValues}). The expected $2/3$-power of $k$ with the wind velocity is nicely consistent with the data, although the limited accessible range of $V$ gives a low sensitivity on the value of the exponent. The adjusted multiplicative factor in front of $(V/v_c)^{2/3}$ is furthermore only $20\%$ below the prediction. The collapse for $\omega$ is less impressive and one observes that the expected proportionality relationship (\ref{expscalingomegaV}) only holds for large velocities: extrapolating the data, $\omega$ would vanish for $V \simeq 1.5 v_c$. Such a multiplicative factor of order one shows that the dimensional analysis of perturbation effects is correct. Similarly, the slope of this linear law is around $2$, which is the correct order of magnitude, but quantitatively larger than the prediction.

\begin{figure}[t]
\centering
\includegraphics{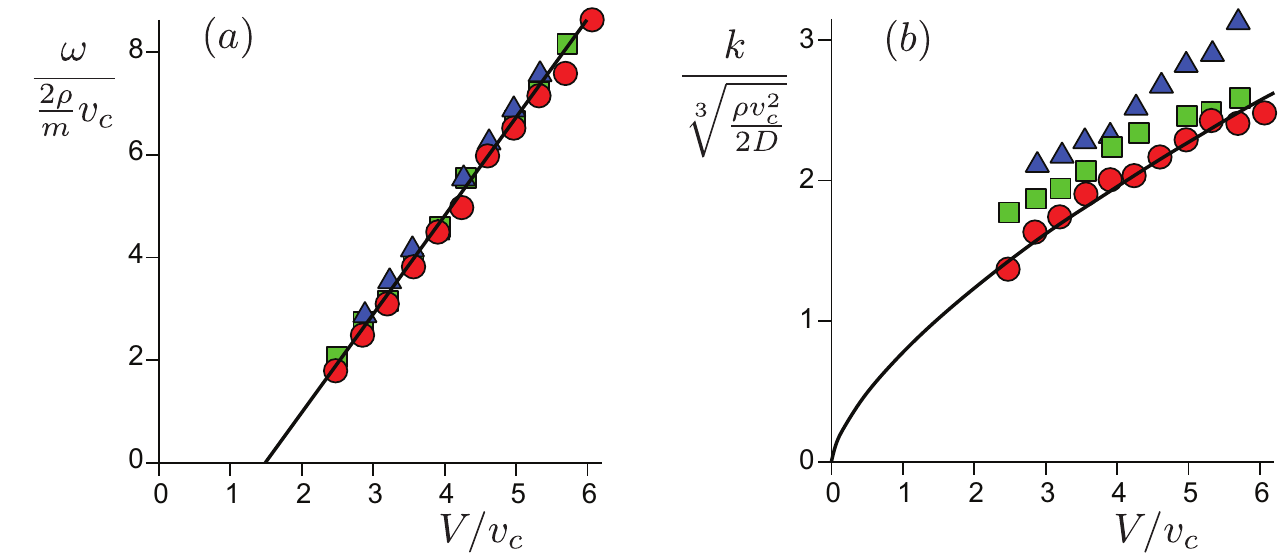}
\caption{Same as Fig.~\ref{ExpScalingsomegamkm}, for the paper material only but with different sheet lengths: $\Delta L = 4$~cm (circles), $\Delta L = 8$~cm (squares) and $\Delta L = 12$~cm (triangles). The solid lines are the same as in Fig.~\ref{ExpScalingsomegamkm} and correspond to the limit of small $\Delta L$ (linear regime).}
\label{omegakDeltaL}
\end{figure}

\begin{figure}[t]
\centering
\includegraphics{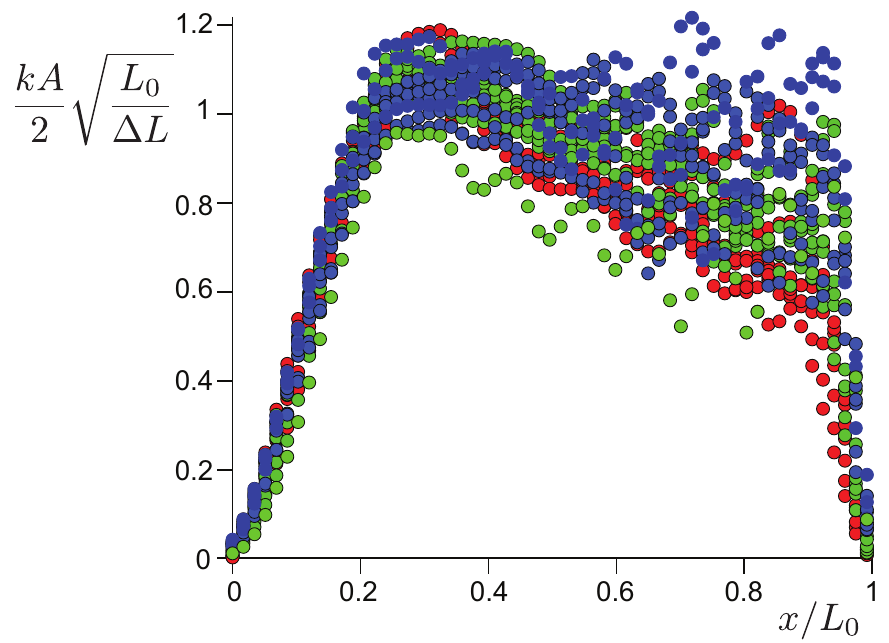}
\caption{Rescaled longitudinal profiles of the wave amplitude. Symbols: red circles ($\Delta L=4$~cm) green circles ($\Delta L=8$~cm) and blue circles ($\Delta L=12$~cm). All these data correspond to the paper sheet. Data for different wind velocities have been gathered.}
\label{AmplitudeDeltaL}
\end{figure}

\subsection{Finite amplitude effects}
Although the results displayed in Fig.~\ref{ExpScalingsomegamkm} show a good agreement of the selection of angular frequency and wavenumber in the experiments with the prediction of the linear stability analysis of the problem, we have also found some evidence for finite amplitude effects. Focusing on the paper material, we have systematically varied the sheet length. Data corresponding to different values of $\Delta L$ are displayed in Fig.~\ref{omegakDeltaL}, showing $\omega$ and $k$ in the same rescaled way as in Fig.~\ref{ExpScalingsomegamkm}. The scaling law obeyed by the angular frequency is found to be independent of the sheet length, whereas that of the wavenumber shows small but systematic variations with $\Delta L$. This shows the presence of non-linearities that are not described here, the linear regime corresponding to the limit of vanishing $\Delta L$.

In fact, wavenumber and amplitude of the waves can be related to each other as follows. Taking a sinusoidal shape $\zeta=A \sin(kx)$ for the sheet over its entire length between the two clamps, the geometrical constraint that the extra-length $\Delta L$ accommodates these undulations without any longitudinal extension can be written as:
\begin{equation}
\Delta L =L-L_0=\int_0^{L_0} (\sqrt{1+\zeta'^2}-1) {\rm d}x.
\label{DeltaLintegral}
\end{equation}
In the regime of small perturbation where $kA \ll 1$, and assuming that $L_0$ is much  larger than the wavelength, the integral $\frac{1}{L_0} \int \sin(kx) {\rm d}x$ vanishes and the above relation can be simplified into:
\begin{equation}
\frac{\Delta L}{L_0} =\frac{k^2A^2}{4} \, .
\label{DeltaL}
\end{equation}
This suggests to take data such as those displayed in Fig.~\ref{BasicData}d, and to produce rescaled amplitude profiles of the form $\sqrt{\frac{L_0}{\Delta L}}\,\frac{kA(x)}{2}$ \emph{vs} $\frac{x}{L_0}$. This is done in Fig.~\ref{AmplitudeDeltaL}, with data corresponding to different $\Delta L$ and different wind velocities. Although scattered, the data collapse and the order of magnitude indicate that this geometrical constraints capture the finite amplitude selection of the waves in these experiments.

\subsection{Conclusion}
Combining experiments and a linear analysis in the study of propagative waves on a flexible sheet submitted to a permanent wind, we have achieved a data collapse along scaling laws showing the selection of their frequency and wavenumber. The experiments have been performed with different materials, varying their rigidity by two orders of magnitude. We have shown that these laws result from the balance between dynamic pressure and inertia or elasticity. However, we have here performed the simplest theoretical analysis, based on an unbounded homogeneous sheet. As a consequence, the theory can only work in the limit $kL_0 \gg 1$. The boundary conditions actually break the invariance along the $x$-axis. In principle, one should therefore find the temporal modes whose spatial shape is a superposition of modes, characterized by a complex wavenumber $K=k-iq$, and satisfying the four boundary conditions -- these are $\zeta(0,t)=0$, $\zeta' (0,t)=0$, $\zeta(L_0,t)=0$ and $\zeta' (L_0,t)=0$. As seen in Fig.~\ref{AmplitudeDeltaL}, the wave amplitude $A$ vanishes at both ends while the central part remains almost homogeneous. We therefore expect the normal mode of the problem to be close enough to a Fourier mode to justify the assumption made here.

Finally, this work can be of interest for applications in energy harvesting. It is generally based on a fluid flow, or on surface waves inducing a relative motion between articulated parts, which is then converted into an electrical current. However, these moving parts are subject to mechanical wear and may be noisy. It would therefore be interesting to use instead deformable systems without rotating parts, like those investigated here: using the steady relative flow between the device and the surrounding fluid to produce energy\cite{ ProcRSocA-Singh2012,APL-Erturk2010,JFM-Michelin2013,JFS-Doaré2011,JSV-Pineirua2015,PhysRevApplied-Xia2015,POF-Akcabay2012}. Any hydrodynamic flow deforming a soft surface on which electrical charges are deposited will lead to a motion of charges, and therefore to a current. Such a soft system therefore enables to transform mechanical energy into electrical energy. The next step is to design a prototype of energy harvesting device based on this principle.

\begin{acknowledgments}
We are grateful to J. Bico and B. Roman for their help in the measurement of the bending rigidity. We thank the Chinese Scholarship Council and Xi'an Jiaotong University for funding (NSFC11402190). 
\end{acknowledgments}

%

\end{document}